\documentclass[a4paper,nocompress]{spie}  
\usepackage[]{graphicx}

\title{Integrated optics interferometric four telescopes nuller} 

\author{Ronny Errmann\supit{a,b}, Stefano Minardi\supit{a}, Lucas Labadie\supit{c}, \\Felix Dreisow\supit{a}, Stefan Nolte\supit{a}, and Thomas Pertsch\supit{a} 
\skiplinehalf
\supit{a}Institute of Applied Physics, Abbe Center of Photonics, Friedrich-Schiller-Universit\"at Jena, Max-Wien-Platz 1, 07743 Jena, Germany; \\
\supit{b}Astrophysical Institute and University Observatory, Friedrich-Schiller-Universit\"at Jena, Schillerg\"a{\ss}chen 2-3, 07745 Jena, Germany; \\
\supit{c}Physikalisches Institut, Universit\"at zu K\"oln, Z\"ulpicherstr. 77, 50937 K\"oln, Germany
}

\authorinfo{Further author information: S.M.: E-mail: stefano@stefanominardi.eu, Telephone: +49 (0)3641 947 848\\  R.E.: E-mail: Ronny.Errmann@uni-jena.de, Telephone: +49 (0)3641 947 518}

\begin{document} 
  \maketitle 

\begin{abstract}
Nulling interferometry has been identified as a competitive technique for the detection of extrasolar planets. The technique consists in combining out-of-phase pairs of telescopes to null effectively the light of a bright star an reveal the dim glow of the companion. We have manufactured and tested with monochromatic light an integrated optics component which combines a linear array of 4 telescopes in the nulling mode envisaged by Angel\&Wolf\cite{ang97}. Our testbench simulates the motion of a star in the sky. The tests have demonstrated a nulling scaling as the fourth power of the baseline delay.
\end{abstract}



\section{INTRODUCTION}
\label{sec:intro}  

Nulling interferometry\cite{bra78} has been for long identified as a unique and competitive technique to perform mid-infrared spatially resolved spectroscopy of Earths and Super-Earths orbiting nearby main-sequence stars\cite{lab14}.  The original Bracewell setup consisted in phase-shifting by 180 degrees one of the beam of a 2-telescope Michelson interferometer in order to cancel out the light of the on-axis bright star and reveal the dim glow of the off-axis companion. By combining in nulling mode more than two telescopes, a broader central dark fringe can generally be obtained\cite{men97}, which allows us to further reduce the stellar leaks resulting from the star being resolved by the interferometer, and furthermore to modulate the exo-zodiacal signal. In addition, a broader null allows higher instrumental tolerances with respect to the residual piston phase and photometric unbalance. Infrared space-based interferometry has been historically the preferential way to implement the nulling technique since optimal sensitivity and spectral coverage in the mid-infrared can be guaranteed from space to detect the photons from the faint planet. The technique has benefited from numerous technical and feasibility studies to investigate the requirements for space operation\cite{coc09}. The inherent complexity of the beam combination stage in a multi-aperture interferometer has always been a pitfall for nulling, and for space interferometry in general. It has been often suggested that the integrated optics (IO) approach would be able to highly simplify the optical design of such an instrument\cite{lab08}, but was experimentally tested only at the level of a simple two-aperture Bracewell setup\cite{web04}, which nevertheless mitigates the advantage of the multi-aperture interferometry approach. While four-beam interferometric nulling concepts have been successfully studied at JPL with the Planet Detection Testbed\cite{mar12}, the potential of the IO for implementing a four-aperture - or plus - design has hardly been investigated both theoretically and experimentally.

Here we concentrate on exploring a four-beam combination nulling scheme based on cascaded 2x2 directional couplers, which have inherent $\pi$-phase shifted outputs. Combining together the nulled outputs is the condition for producing the necessary broad dark fringe. In this paper, the accent is set on showing the proof-of-concept of an integrated optics four-telescope nuller rather than on achieving and optimizing deep level of nulling. We therefore report, for the first time to our knowledge, the characterization of an integrated optical component designed to combine in nulling mode 4 telescopes. Our investigation has been conducted at visible wavelength (632\,nm), but the emergence of new photonics solutions operational in the 10-microns spectral range\cite{lab09} justifies the interest of our approach in preparation of future mid-infrared instrumentation. 


\section{EXPERIMENTAL SETUP}

\begin{figure}
   \begin{center}
   \begin{tabular}{cc}
   \includegraphics[height=5cm]{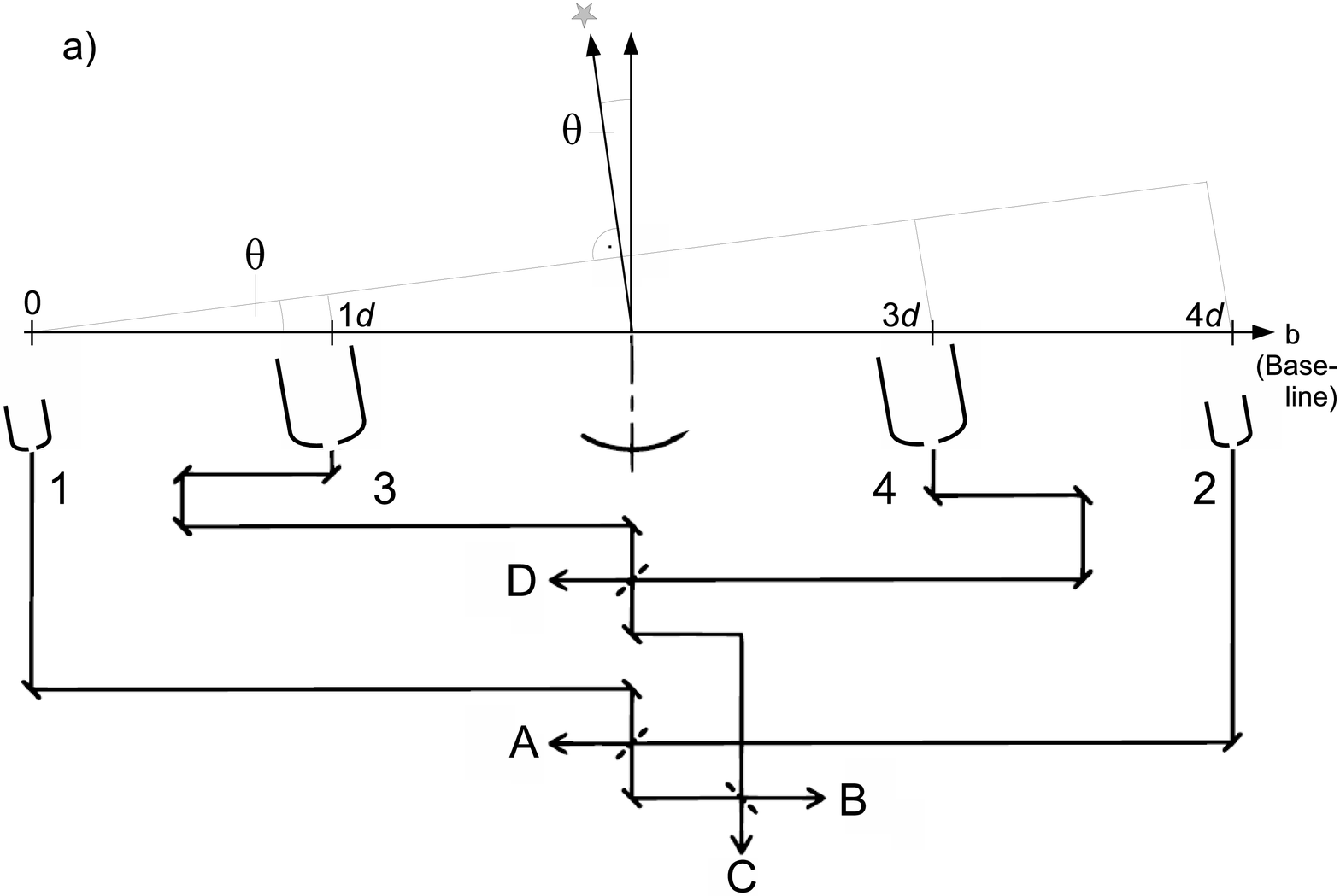}
   &
   \includegraphics[height=5cm]{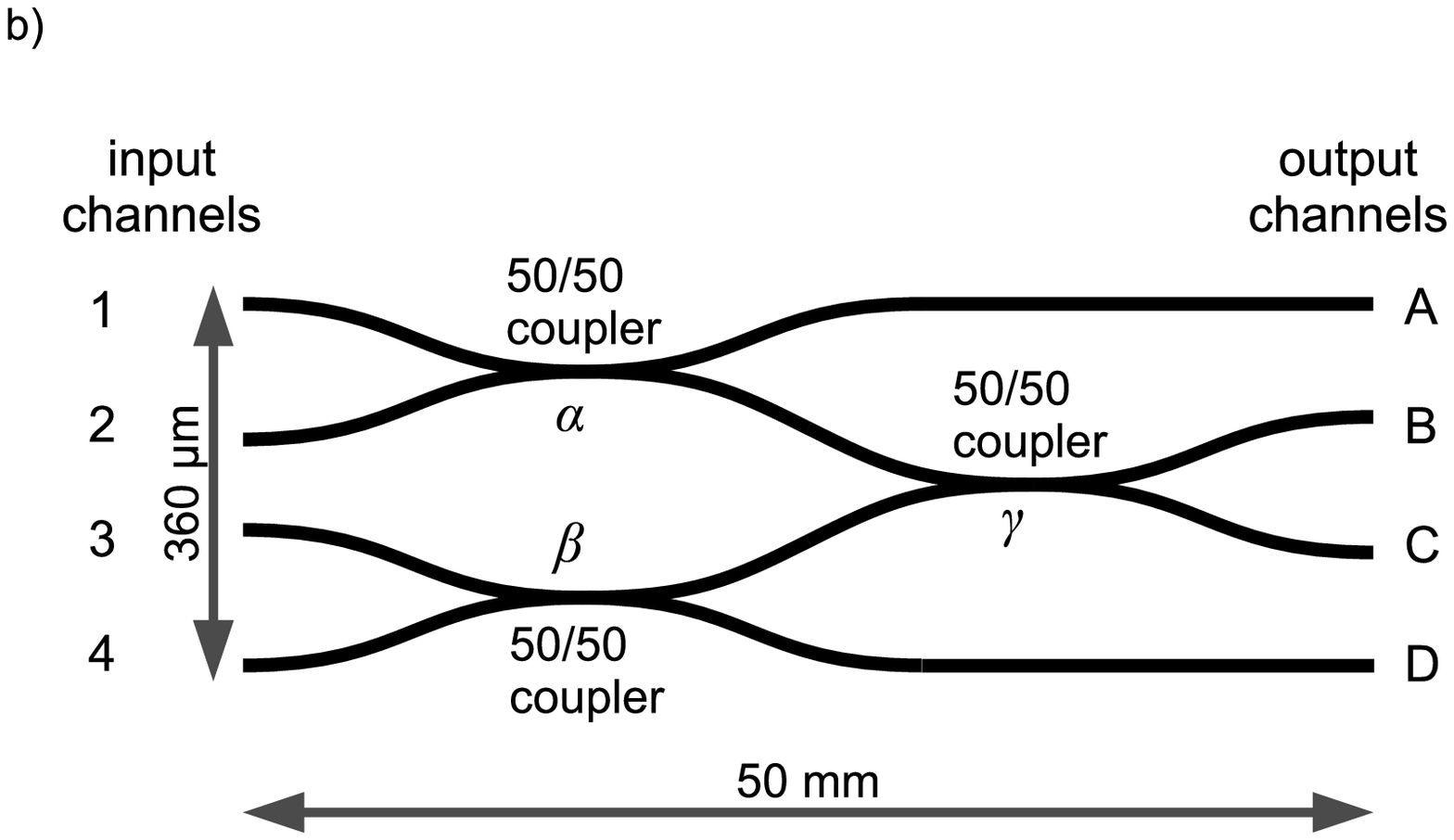}
   \end{tabular}
   \end{center}
   \caption[Telescope setup] 
   { \label{fig:teles} 
Left: Telescope arrangement, for which the nulling interferometry was simulated. The mirror diameter of the telescopes at the edge (telescopes 1 and 2) are half the diameter of the inner ones (telescopes 3 and 4). Right: Setup of the photonic chip, using 3 beam combiners ($\alpha$, $\beta$, $\gamma$) with 50/50 coupling ratio each. The numbers of the input channels correspond to the number of the telescopes.}
   \end{figure}
   
Our goal is to investigate the potential of the integrated optics approach for multi telescope nulling interferometry. To this end, we manufactured and tested an optical chip which reproduces the functionalities of the beam combiner of the classical Angel \& Woolf\cite{ang97} configuration for multi-telescope nulling. In its basic form, this configuration (illustrated in Fig.~\ref{fig:teles}.a) consists in a linear array of 4 telescopes located at coordinate 0,$d$, $3d$ and $4d$, respectively. The two inner telescopes (3 and 4) collect a flux which is four times the one collected by the two external telescopes (1 and 2). The beam combiner of the interferometer consist of three 50/50 beam splitters arranged on two levels. At the first level, two beam splitters are used to combine pairwise the internal and the external telescopes in nulling mode, \textit{i.e.} the phase delay between the telescope pairs is arranged such that at each beam splitter one output exhibits a destructive interference (nulled output). 
At the second level, a third beam splitter is used to combine the nulled outputs of the previous level of beam combination.

The implementation in terms of integrated optics of the 4-telescopes nulling beam combiner is straightforward. The scheme of the integrated optical circuit is shown in Fig.~\ref{fig:teles}.b. Light propagates from left to right. The beam splitters are replaced by 50/50 waveguide couplers and are arranged in a two-level cascade, the third coupler connecting two of the outputs of the first two. The four input waveguides are connected to the respective telescopes according to the numeration of Fig.~\ref{fig:teles}.a. 
The IO component was fabricated by femtosecond laser inscription in silica substrate\cite{per04} and has the physical dimensions of $360\,\mu$m$\times 50$\,mm, as indicated in the Figure. Waveguides are strictly single mode for the wavelength we have chosen for the experiment ($\lambda=633$\,nm, corresponding to the emission line of a He-Ne laser). 
    

   \begin{figure}[t]
   \begin{center}
   \begin{tabular}{c}
   \includegraphics[width=0.95\textwidth]{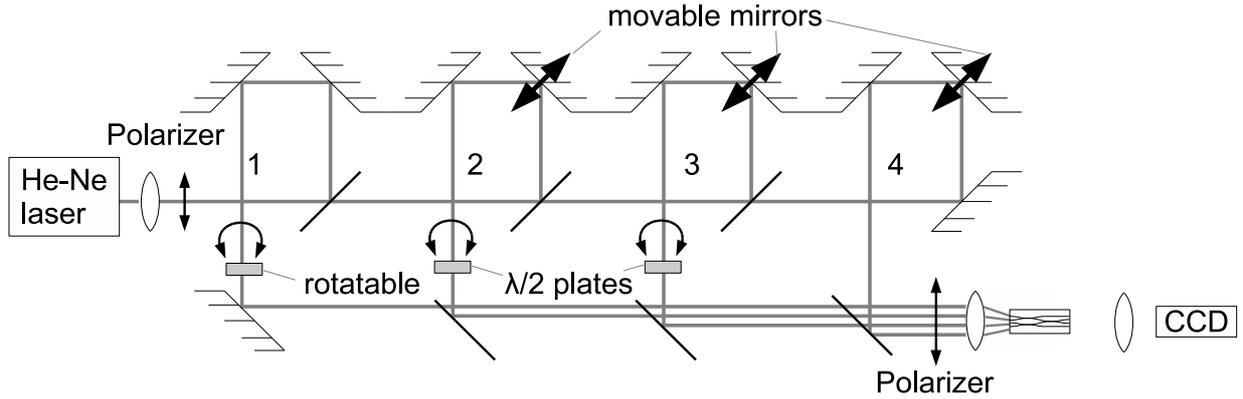}
   \end{tabular}
   \end{center}
   \caption[Setup] 
   { \label{fig:setup_tel} 
Scheme of the setup used to simulate the four telescopes including the phase modulation and flux control. The flux of the beams 1 and 2 are adjusted to 25\% of the beams of 3 and 4. The integrated optics device for the beam combination is shown also in Fig.~\ref{fig:teles}.b.}
   \end{figure}

The simulation of the telescopic array and the motion of the star in the sky required the implementation of a rather complex setup which is schematically outlined in Fig.~\ref{fig:setup_tel}. It consists basically in a modified Mach-Zehnder interferometer which is used to prepare 4 beams out of a collimated and astigmatically shaped laser beam (as mentioned, we used a He-Ne laser as a source). The four beams are recombined with a small tilt to each other in the horizontal direction so that they can be focused at the input of the corresponding waveguides of the photonic chip by means of a NA=0.25 microscope objective (the numbering of the beams in Fig.~\ref{fig:setup_tel} corresponds to the telescopes/waveguides illustrated in Fig.~\ref{fig:teles}).
Astigmatic beam shaping of the laser beam was necessary to improve the coupling efficiency into the laser-written waveguides of the photonic chip, which support elliptical propagating modes\cite{min12}.

The brightnesses of the different beams is controlled by a combination of rotatable $\lambda$/2 plates and a polarizer. As expected for the Angel\&Wolf configuration, the input beams 1 and 2 are set to a quarter of the flux of the beams 3 and 4. In order to include the effect of the coupling efficiency in the waveguides (which we did not measure), this calibration was performed by measuring the total flux at the output of waveguides ABCD for single beam injection. The measurement of the output fluxes (described in the next section) was carried out on images of the output facet of the photonic chip recorded with a high-resolution, 10-bit CCD camera attached to a microscope. 

To simulate the motion of a target star in our model interferometer, we control the relative optical path difference (OPD) between the 4 beams with three movable mirrors, which we fabricated by attaching the mirrors to the membrane of loudspeakers. 
If the star transits near the zenith, the $\mathrm{OPD}$ measured at each telescope respect to the first one is given by $\mathrm{OPD_n}=b_\mathrm{n}\cdot\theta$, where $b_\mathrm{n}$ is the baseline between the first and the $\mathrm{n^{th}}$ telescope (n=2,3,4) and $\theta$ is the zenital angle (see Fig.~\ref{fig:teles}.a).  
In our setup, we simulate the linear dependence of the OPD on the angle $\theta$ by feeding the loudspeakers with a sawtooth driving voltage whose amplitude is proportional to the baseline.
Therefore, we kept fixed the path of telescope 1 (our reference) and modulated the delay for the beams 2, 3 and 4 with a signal of OPD amplitude  $2\,\lambda$, $0.5\,\lambda$, and $1.5\,\lambda$ (corresponding to baseline $4d$, $1d$, and $3d$), respectively. We chose to map the baseline $d$ on an amplitude of $0.5\,\lambda$ as we are limited by the linear response of the loudspeakers in respect to the input voltage. 
The phase delay control of the setup as well as the recording of the output of the IO are automatized with a \texttt{LabVIEW} program.

\section{CHARACTERISATION OF THE PHOTONIC CHIP}

 \begin{figure}[b]
   \begin{center}
   \begin{tabular}{c c}
   \includegraphics[width=0.45\textwidth]{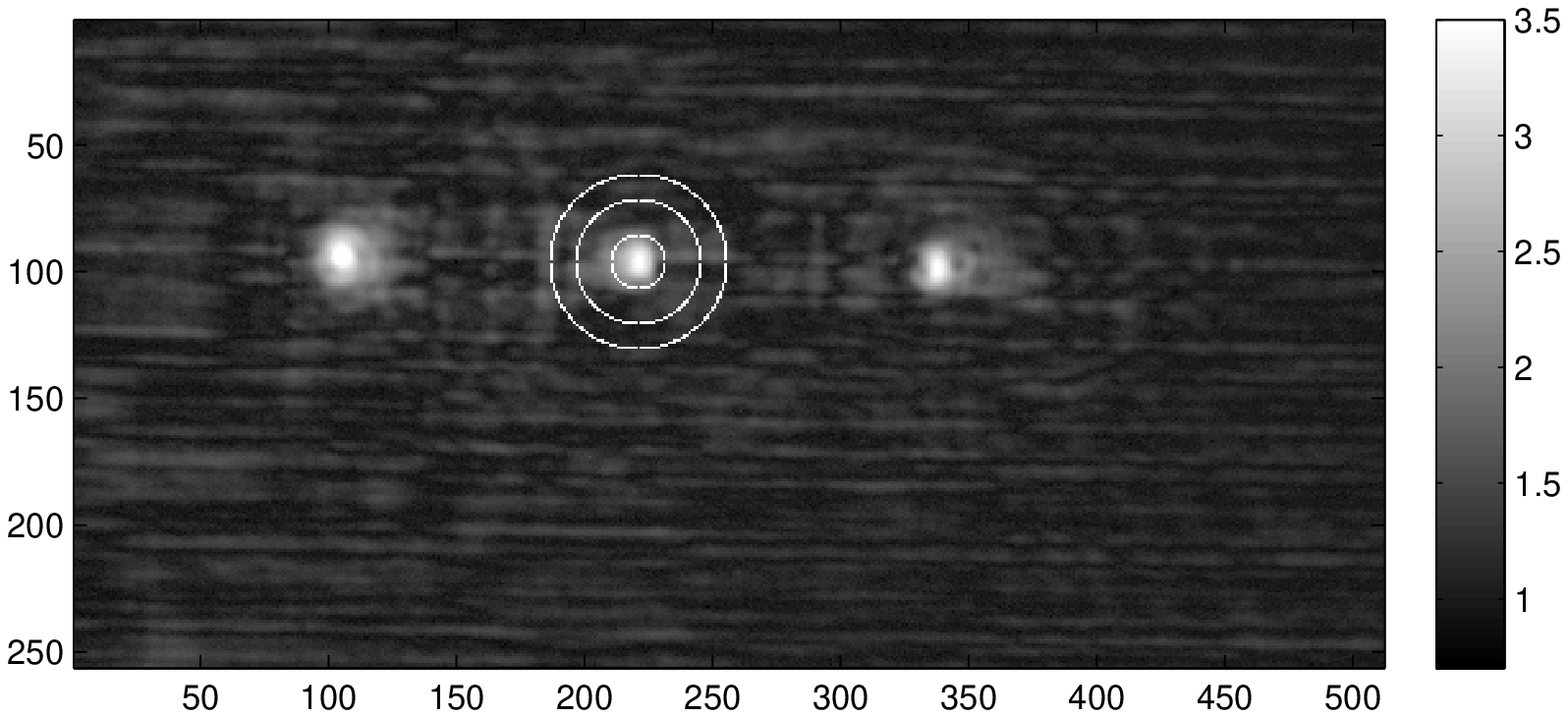} & \includegraphics[width=0.45\textwidth]{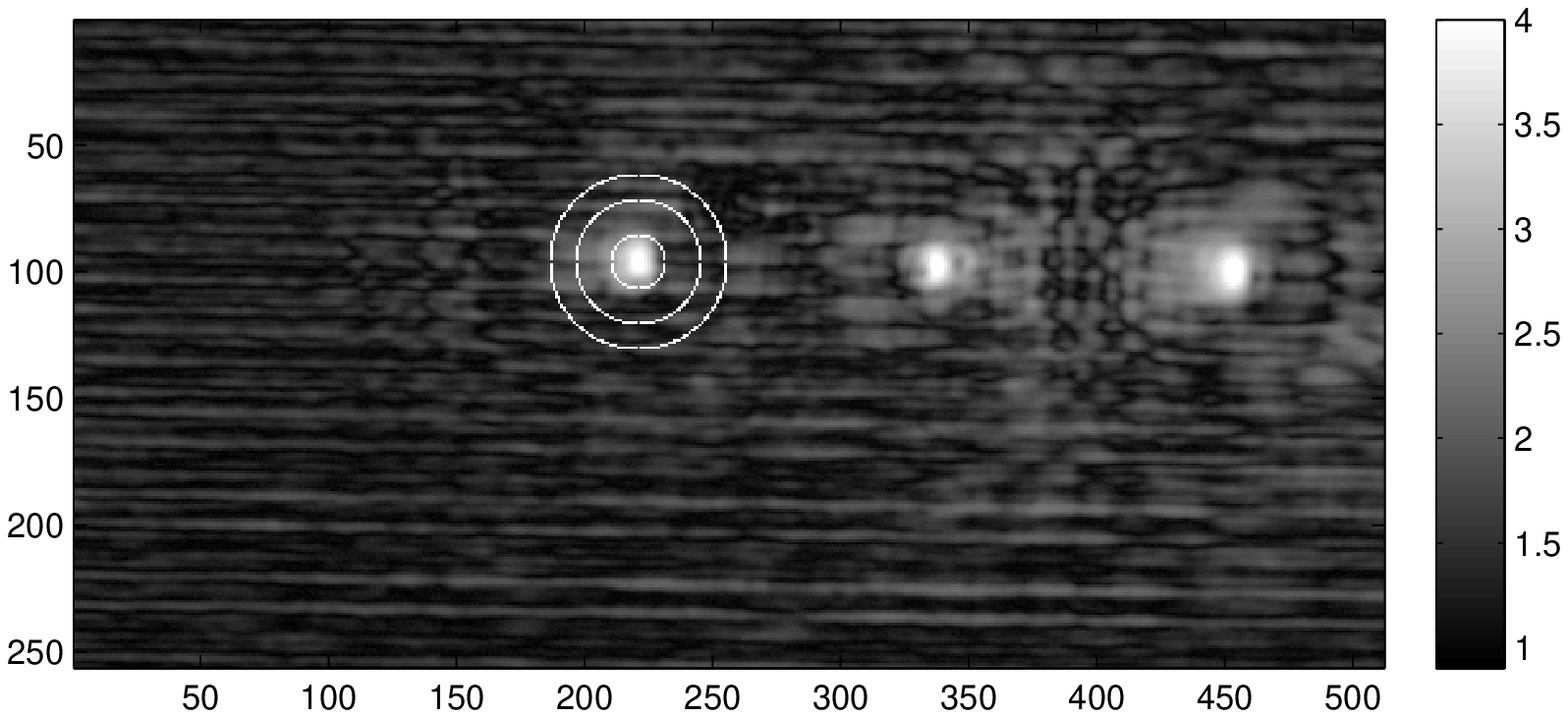} 
   \end{tabular}
   \end{center}
   \caption[Output field] 
   { \label{fig:output_field} 
Logarithmic output field for the channels A, B, C, D. Light was coupled into the input channels 1 (left image) and 4 (right image). The aperture (10\,px radius, corresponding to $10.2\,\mu$m), which includes 90\% of the light, as well es ring shaped area for the background measurements (25-35\,px) are shown in the images. Grayscale in decadic logarithm of the counts.}
   \end{figure} 

The characterization of the laser-written photonic chip consisted in measuring the splitting ratio of the couplers.
As mentioned in Section 2, the 4-telescope nuller requires beam splitting ratios of exactly 50\%. 
In order to cope with the fabrication uncertainties associated with laser writing, several samples of the beam combiner were written in a glass substrate and tested. Samples differ from each other by the separation of the waveguides at the center of the couplers, which controls the coupling ratio between the waveguides. Samples with waveguide separations between 17 and $19\,\mu$m in steps of $0.5\,\mu$m were written and characterized individually. 

The splitting ratios were inferred from the photometry of the output waveguides ABCD for single waveguide excitation of the photonic chip.
We employed the aperture photometry method, that is measuring on CCD images the flux inside a circular aperture around the center of an output waveguide and correct it with the background flux. The latter one is calculated in a ring around the aperture with a gap between ring and aperture. We used a radius of 10\,pixel (px) for the aperture, which corresponds to $10.2\,\mu$m at the photonic device, and a ring starting at a distance 25\,px from the center and a width of 10\,px. 
Fig.~\ref{fig:output_field} shows two typical images from the output facet of the photonic chip and overlaid the photometric region of interest corresponding of output waveguide B. 
With this method, we can estimate the light power carried by each waveguide, regardless of distortions of the mode shape associated to 
interfering stray light (estimated to be about 1\% of the peak intensity of the mode). 

The properties of the best integrated beam combiner are given in Table~\ref{tab:beam_comb}. 
In the left part of the table, we give the values of the normalized photometric outputs for waveguides A-D (columns) for individual input waveguide excitation (1-4, lines). Subsequently, the splitting ratios of the couplers were calculated (right part of the table) under the assumption that losses from all connecting waveguides are negligible. We reach nearly 50\% beam splitting ratios for all 3 beam splitters, the largest difference to that is only 2.5\%. Interestingly, the values in each of columns B and C differ slightly, depending on the injected input channel. The largest difference occurs at output B when switching between inputs 1 and 2. As the light passes the same paths after the beam splitter $\alpha$, one would expect the same flux ratios between output B and C. The difference is most probably due to an interference between coherent stray light, which differs between different excitation of the waveguides, and the guided signal in the waveguides.

\begin{table}[h]
\caption{Normalized output flux for single input injection and the calculated beam splitter efficiencies. See Fig.~\ref{fig:teles} for the definition of the variables, the flux ratio is given for the upper waveguide. The ratio and uncertainty of the splitter efficiency are the average and standard deviation of the output channels, respectively.} 
\label{tab:beam_comb}
\begin{center}       
\begin{tabular}{|r|cccc|l| @{\hspace{10mm}}   rcl  |}
\hline
Input   & \multicolumn{4}{c|}{Output channels}        &        & \multicolumn{3}{c|}{beam splitter ratios}   \\
channel & A        &     B      &   C     &   D       & sum    & beam splitter & ratio & uncertainty         \\
\hline
1       & 0.495    & 0.258      & 0.247    & 0.001    & 1.001  & $\alpha$ & 0.503 & 0.011          \\ 
2       & 0.511    & 0.224      & 0.264    & 0.001    & 1.000  & $\beta$  & 0.476 & 0.004          \\ 
3       & 0.002    & 0.274      & 0.250    & 0.474    & 1.000  & $\gamma$ & 0.502 & 0.030          \\ 
4       & 0.000    & 0.270      & 0.251    & 0.479    & 1.000  &          &       &                \\
\hline 
\end{tabular}
\end{center}
\end{table} 

\section{INTERFEROMETRIC RESULTS}

\begin{figure}
   \begin{center}
   \begin{tabular}{cc}
   \includegraphics[width=0.48\textwidth]{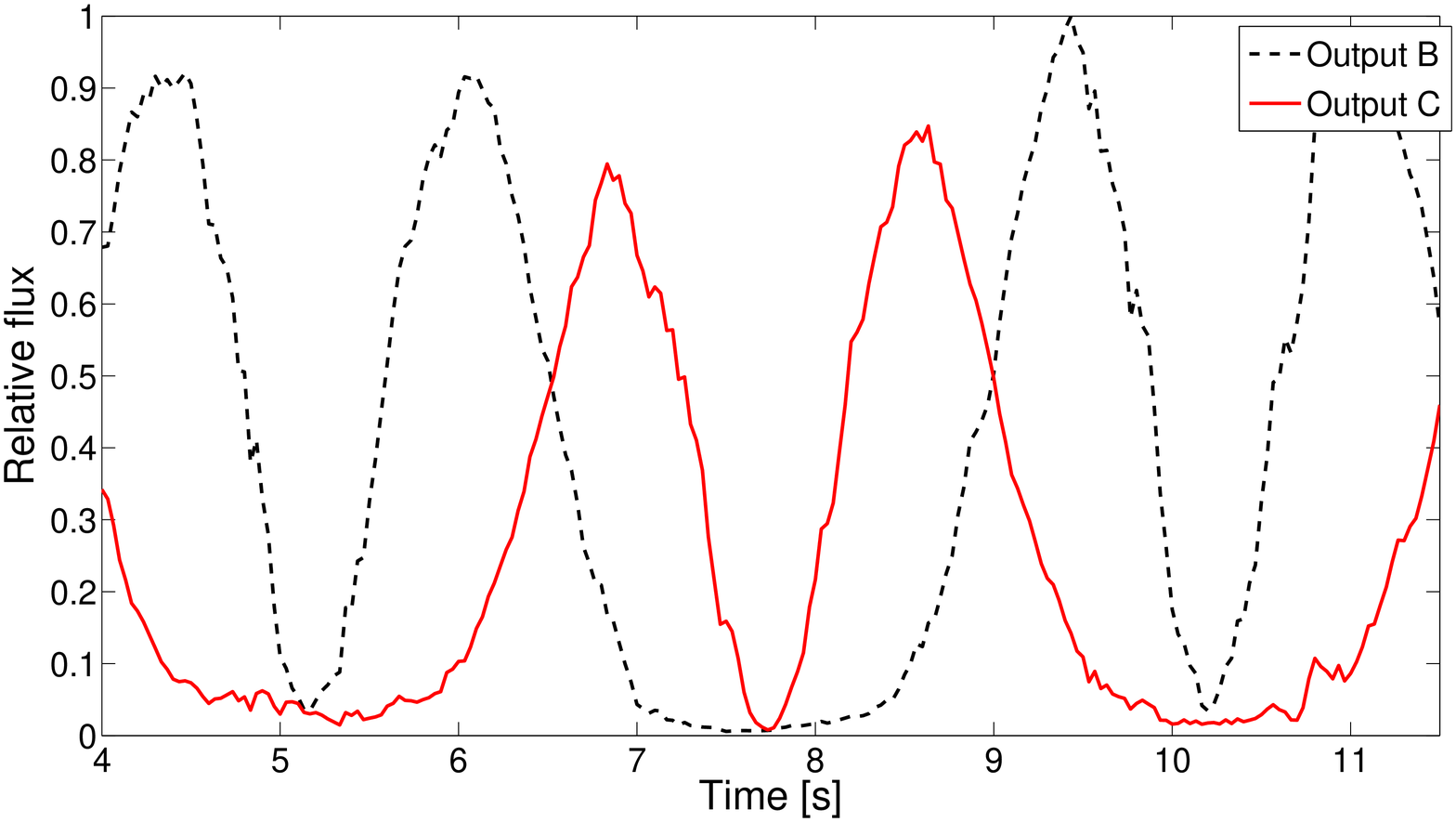}
   &
   \includegraphics[width=0.48\textwidth]{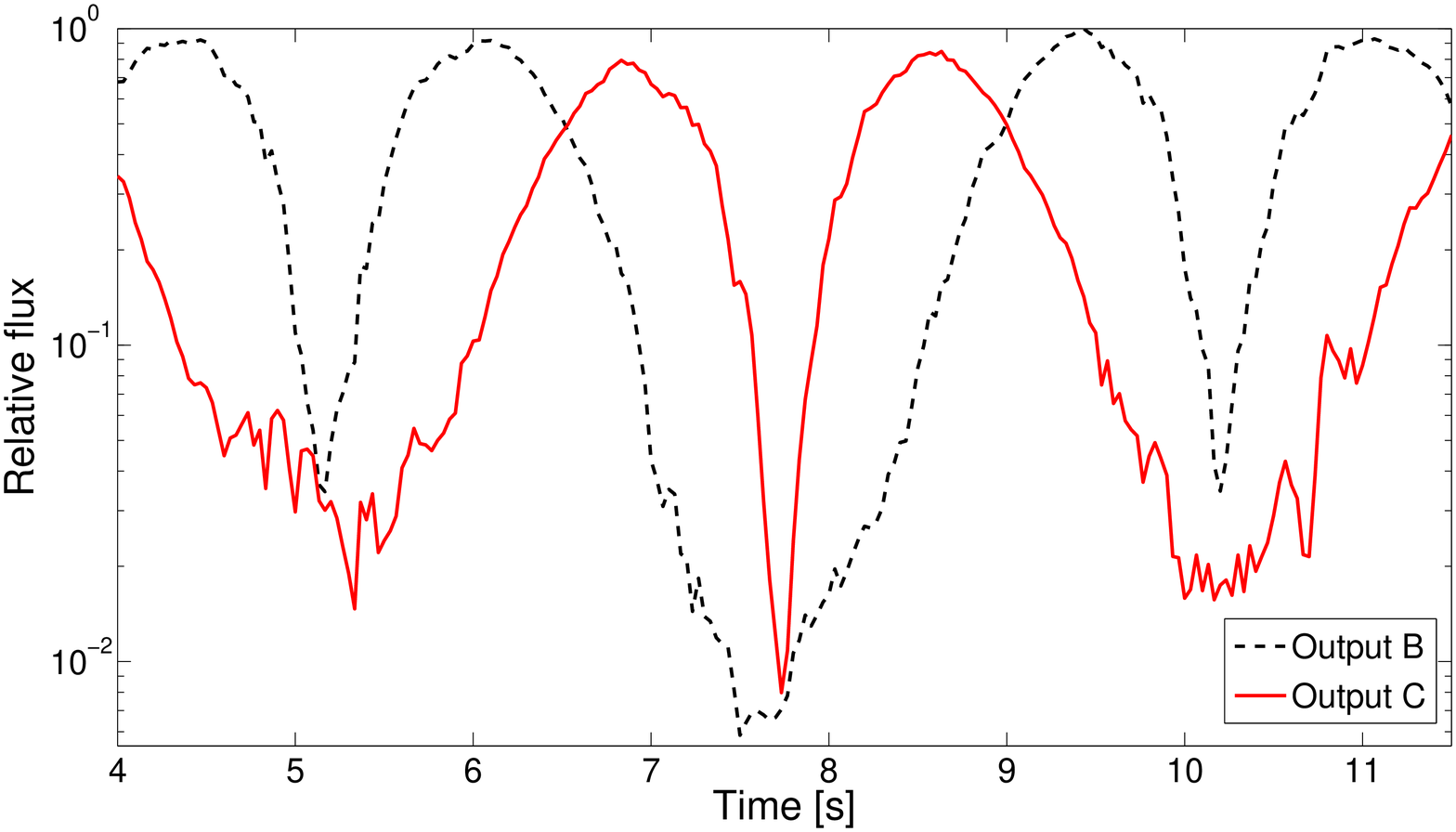}
   \end{tabular}
   \end{center}
   \caption[Setup] 
   { \label{fig:outBC} 
Output B and C as a function of time, in linear (left) and logarithmic (right) scale. A smoothing over 3 sampling points (0.1 s) is applied to the data to reduce noise.}
   \end{figure}

The interferometric trace of the deep nulling outputs B and C is illustrated in Fig.~\ref{fig:outBC}.a.
As the angular coordinate advances (in our case, the time), we have an alternating sequence of  narrow and broad minima for each of the outputs. 
The logarithmic plot shows that the depth of the nulling signal ranges form 1:30 to 1:100. 
This signal sequence is expected from the theory of the 4-telescope nulling. For applications to astronomy, the broad minima are the ones of interest, since they can mitigate the stellar leakage and the phase delay errors which would occur in a real world interferometer.  
In fact, Angel \& Wolf predict that for small angular detunings $\theta$ from the center of the broad null the 4-wave interference signal should scale as $\theta^6$, thus suppressing efficiently the light over a large range of angles.

\begin{figure}[b]
   \begin{center}
   \begin{tabular}{c}
   \includegraphics[width=0.95\textwidth]{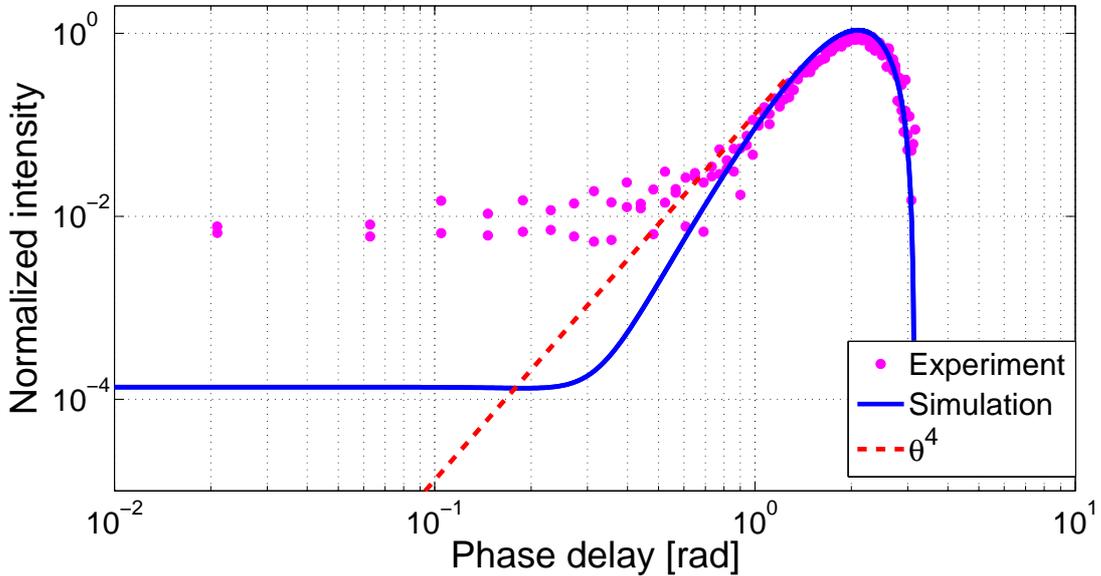}
   \end{tabular}
   \end{center}
   \caption 
   { \label{fig:loglogplot} Bi-logarithmic plot showing the power dependence of the interferometric signal as a function of the detuning from the center of a broad minimum. The high background level prevents to observe the $\theta^6$ dependence of the interferometric signal. Overlaid in full line, the simulated fringe from an imperfect nulling beam combiner featuring the splitting ratios we measured in the chosen combiner.}
   \end{figure}

To test the power law describing the angular dependence of the interferometric signal in proximity of the center of the broad minimum, we plot the output signal of channel B as a function of the absolute value of the phase detuning (defined as $\phi=k\,\theta\,d$, $k$ being the wavenumber of light)  in a bi-logarithmic plot (see Fig.~\ref{fig:loglogplot}). We found that for delays larger than 0.5 rad the interferometric signal follows a $\theta^4$ power law. For smaller delays, the interferometric signal is relaxing to a constant value of about 1:100. Simulations show that for a perfect combiner the $\theta^6$ behavior should become apparent for phase detunings smaller than 0.2-0.3 rad. At that point, however, the nulling depth would already reach levels below $10^{-4}$, requiring a detection system with a much higher dynamic range than the employed 10-bit CCD.

\section{Discussion}
Even considering the reduced dynamics of our measurement system, the background level observed in the experiments is significantly higher than the noise floor, estimated in about 0.1-0.2\%. To account for this observation, there are a few possible causes. From one side, we have seen (see Tab.1) that the splitting ratio of the couplers is not the same and coupler $\beta$ deviates by nearly 3\% from the ideal value of 50\% splitting ratio. 
Simulations show that fabrication errors that lead to a variation of the splitting ratio of the couplers can result indeed in a constant noise floor that can reduce the contrast of the fringes. One of such simulations featuring the measured coupling ratios is illustrated in Fig.~\ref{fig:loglogplot}. We see the appearance of a constant floor for $\phi<0.1$ rad, albeit with a much deeper null than observed in the experiment. 

A second source of noise comes from the stray light which is generated by 1) the non-perfect coupling of the light into the waveguides due to mode mismatch and 2) radiation losses from the bended waveguides. 
As seen in the logarithmic intensity maps displayed in Fig.~\ref{fig:output_field}, these two sources of stray light have an intensity of about 0.1\% the peak value of the guided mode intensity distribution. While the incoherent part of this background is taken into account by the employed photometric method (see Section 3), it is not possible to remove its coherent contribution to the signal. 
Better control of this stray light may be accomplished by bending the input/output waveguides so that the injection light beam direction is not collinear to the observation direction, or by implanting suitable slits in the chip. Both methods were successfully used in the past to achieve deep nulling levels with integrated optics\cite{web04}. 

Finally, we mention that other sources of noise may come from small photomertric and phase variations which prevent the achievement of a deep nulling. Indeed, small displacements of the beam induced by the phase modulators could alter the  coupling ratio in the waveguide by a fraction of percent, thus preventing exact nulling at the waveguide couplers.
Also the fact that we do not control with a feedback-loop the optical path difference between the beams 
could play a role. Previous characterization of a similar Mach-Zehnder interferometer installed in our lab, showed that the r.m.s. phase noise is about 0.08 rad\cite{min12}, or 8 nm in terms of OPD.
Better phase control could be achieved by using light sources of longer wavelength and/or a feedback-loop control of the OPD based on laser metrology\cite{mar12}.

\section{Conclusions}

We have characterized for the first time a 4-telescope integrated optics beam combiner for nulling interferometry. 
In our setup we simulated the operation of a linear array of telescopes, as proposed by Angel\&Wolf \cite{ang97} in the context of exoplanet detection.
We showed that broad (in terms of the zenithal angle $\theta$) nulls can be generated, which could in principle mitigate the stellar leakage of partially resolved stars and/or exo-zodiacal light. While our experiments were primarily focused on the proof-of-concept of the IO component, we achieved a nulling depth from 1:30 to 1:100. Deeper nulling was prevented by a number of technical issues (some of them related to the test bench stability) which we will address in future experiments. 

The importance of our work resides in demonstrating the potential of integrated optics for space-based, multi-telescope nulling interferometry. In fact, the use of integrated optical technology removes several degrees of freedom (alignment and positioning) which are present in bulk-optics setups\cite{mar12}, sensibly reducing the complexity and maintenance procedures of the beam combiner.  This is of particular importance for space missions where automated alignment procedures are costly and subject to failure.
We highlight that the chosen manufacturing platform (direct laser writing) is suitable also for the production of integrated components in mid-infrared \cite{rod12}, where exo-planet detection is most favored.

\acknowledgments     
 
RE would like to thank DFG for support in the Priority Programme SPP 1385 in project NE 515 / 34-1 and the Abbe School of photonics for the Ph.D. grant.


\bibliography{report}   
\bibliographystyle{spiebib}   

\end{document}